 \documentclass[reprint]{JASA}

\usepackage{multirow}
\usepackage{soul}
\usepackage{booktabs}
\usepackage{adjustbox}
\sethlcolor{yellow} 

\begin{document}

\title{Depression Markers in Speech: An Approach based on Tract Variables Dynamics}


\author{Sahar Altalhi}
\affiliation{University of Glasgow (United Kingdom) and Taif University (Saudi Arabia)}
\author{Tanaya Guha}
\author{Alessandro Vinciarelli}


\affiliation{University of Glasgow (United Kingdom)}






\preprint{Altalhi et al., JASA}		


\begin{abstract}
This study identifies new depression biomarkers based on the dynamical properties of tract variables, which represent geometric features describing the configuration of the speech articulators. A key advantage of this approach lies in its ability to quantify aspects of the articulatory process that have not been previously explored in the context of depression, namely predictability, complexity, and randomness. These properties are respectively characterised using the Largest Lyapunov Exponent, the Correlation Dimension, and the Sample Entropy. Thorough experiments were conducted on the Androids Corpus, a publicly available dataset comprising 64 speakers diagnosed with depression by clinicians and 54 control speakers with no reported history of mental health conditions. The results indicate that the proposed biomarkers effectively discriminate between the depressed and control speakers, as evidenced by the high Cliff's delta values across both read and spontaneous speech.
\end{abstract}


\maketitle
\section{Introduction}
A long-standing issue in diagnosing mental health conditions is that psychiatrists ``continue to build on subjective clinical assessment'', and the community has often highlighted the need to develop biomarkers that can provide objective measurement opportunities~\citep{Thibaut2018}. In this context, the term \emph{biomarker} refers to any measurable indicator of the severity or intensity of a mental health condition that offers insight into its underlying mechanism and prognosis~\cite{Majkic2011}. This article addresses this knowledge gap by identifying a set of speech biomarkers related to the Major Depressive Disorder (commonly known as Depression) that, to the best of our knowledge, has not been investigated before. 

According to the latest data\footnote{\url{https://www.who.int/news-room/fact-sheets/detail/depression}} from the World Health Organization, more than 4\% of the world’s population experiences depression.  Prevalence is particularly high among adults, affecting 5.7\% globally (4.6\% of men and 6.9\% of women), and rises further to 5.9\% among adults aged 70 years and older. These figures are concerning, and yet they are likely to be an underestimate of the actual numbers, because depression remains undiagnosed (especially in men) due to social stigma~\cite{Covello2020} and lack of suitable healthcare support in large parts of the world~\cite{Wang2007}. Various behavioural cues, both verbal and non-verbal, have been studied in the context of depression; for example, facial expressions~\cite{Girard2013}, head motion patterns~\cite{Gahalawat2023} and lexical choices~\cite{Rude2004}. What makes speech an attractive modality is its availability and effectiveness in detecting depression. High-quality spoken data can be collected easily and unobtrusively with common devices like smartphones and laptops. In addition, only a few seconds of data are shown to be sufficient for detecting this condition with fairly high accuracy~\cite{Aloshban2020}. Since patients with depression tend to experience interactions negatively, mostly due to social impairments typically associated with the pathology, quick and easy ways to detect depression are critical~\cite{Kupferberg2023}. Last, but not least, it is since the earliest times of modern psychiatry that clinicians have observed the effect of depression on speech: `the patients speak in a low voice, slowly, hesitatingly, monotonously, sometimes stuttering'~\cite{Kraepelin1921}. Therefore, speech-based biomarkers, among all other behavioural biomarkers, are both reliable and readily available.

A key assumption underlying the current work is that the \emph{speech articulators}, i.e., the anatomical elements shaping the sound produced during speech, constitute a dynamical system. This system can be understood as a physical system that changes its state over time, where the time dependence is measurable via its articulatory space. Therefore, the \emph{Tract Variables} (TVs), i.e., the geometric features that represent the positions of the speech articulators, can be thought of as the system's \emph{observables}; where these observables are the measurable traces of the articulatory space. Mental health conditions such as depression are known to change the speech production process, which in turn changes the dynamics of the TVs \cite{Williamson2019}. Therefore, a dynamical system-perspective enables analysing the articulatory process in terms of its \emph{predictability} (measured using the Largest Lyapunov Exponent (LLE)) and \emph{complexity} (measured using the Correlation Dimension (CD)). The dynamical system-based measures are complemented with an information-theoretic measure of \emph{randomness}, called the Sample Entropy (SE), that measures the rate at which new information is generated. To the best of our knowledge, these properties of speech production dynamics have not been investigated before in the context of depression detection. These measures are hypothesized to contain depression-relevant information because the `neurophysiological change in depression generally affects motor coordination, including articulatory control and dynamics'~\cite{Williamson2019}. 

The studies reported in this paper have been conducted on the Androids Corpus~\cite{Tao2023}, which is a recently released large depression detection dataset. This is also the only publicly available corpus that is clinically validated, i.e., its subjects with depression have been actually diagnosed by professional psychiatrists. In contrast, the majority of past work used datasets that cannot be considered as clinical samples as they come from volunteers who have had their depression severity established through self-report questionnaires~\cite{Cummins2023}. Therefore, the Androids Corpus is more likely to be representative of real `depressed speech'. In addition, the corpus includes both \emph{read} and \emph{spontaneous} speech for over 90\% of its speakers. The availability of speech collected under diverse conditions increases the chances of discovering relevant biomarkers.

The rest of this paper is organized as follows: Section~\ref{survey} surveys previous work, 
Section~\ref{biomext} describes the approach for the extraction of the proposed biomarkers,
Section~\ref{expres} presents experiments and results, Section~\ref{Classification Experiments} presents the classification of depressed vs control speakers using the biomarkers proposed in this work, and the final Section~\ref{concl} draws some conclusions.
\section{Survey of Previous Work}\label{survey}
The most common approach to identifying depression markers is to automatically extract features from speech signals and compare them between depressed speakers and a \emph{healthy control} group, i.e., speakers not affected by any pathology (mental or physiological). When the difference is statistically significant, then the feature is considered a biomarker. Extensive surveys of the works based on such approaches are available, e.g., in~\citet{Koops2023} for depression and in~\citet{Low2020} for psychiatric issues in general. 

The marker that has been investigated the most is the \emph{speaking rate}, measured in terms of words or syllables per unit of time. The seminal experiments by~\citet{Cannizzaro2004} showed, for the first time, that depression leads to slower speaking and to increased number and length of pauses. The main limitation of the work is that only 7 speakers were involved. However, the findings were confirmed later using different and more substantial corpora by \citet{Tao2020} and \citet{Cummins2023}.
A wider spectrum of markers was considered by~\citet{Menne2024}, who analyzed 44 speakers and focused on features known to be effective in automatic depression detection. Statistically significant differences between the depressed and healthy speakers were observed for several features, such as those related to pitch and energy.

The experiments by~\citet{France2000} involved 115 speakers and showed that the formants and the power spectral density were the most discriminative features while distinguishing between healthy controls and speakers affected by depression, dysthymia or suicidal tendencies. Jitter (variability of the fundamental period in the speech signal) and shimmer (variations in the amplitude of the vocal cord vibration) appeared to be effective markers in a work that involved 9 speakers~\cite{Vicsi2012}. Another analysis on 10 speakers identified glottal flow spectrum~\cite{Ozdas2004} as a possible biomarker. Finally, the experiments by~\citet{Scherer2016} showed that the frequency range of the vowels uttered by depressed speakers tends to be narrower, especially for what concerns vowels \emph{/i/}, \emph{/a/} and \emph{/u/} (253 speakers in total). 

\begin{figure*}[t]
    \centering
\includegraphics[width=0.45\textwidth]{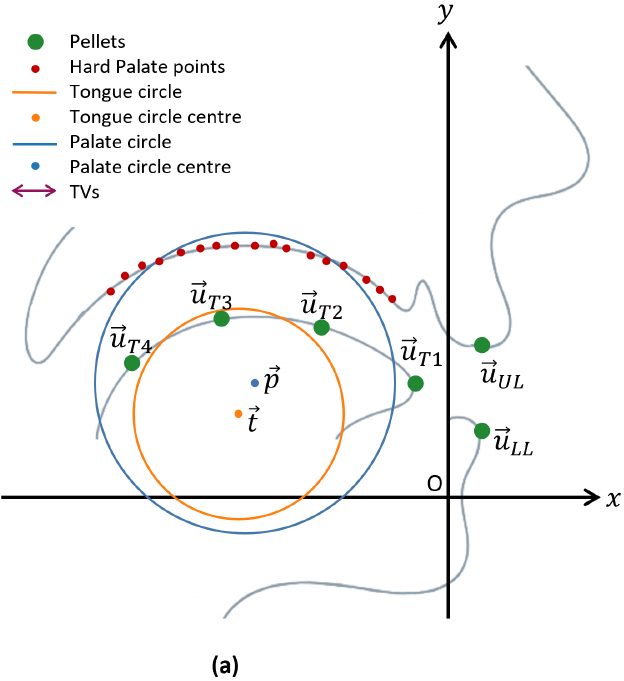}
\includegraphics[width=0.45\linewidth]{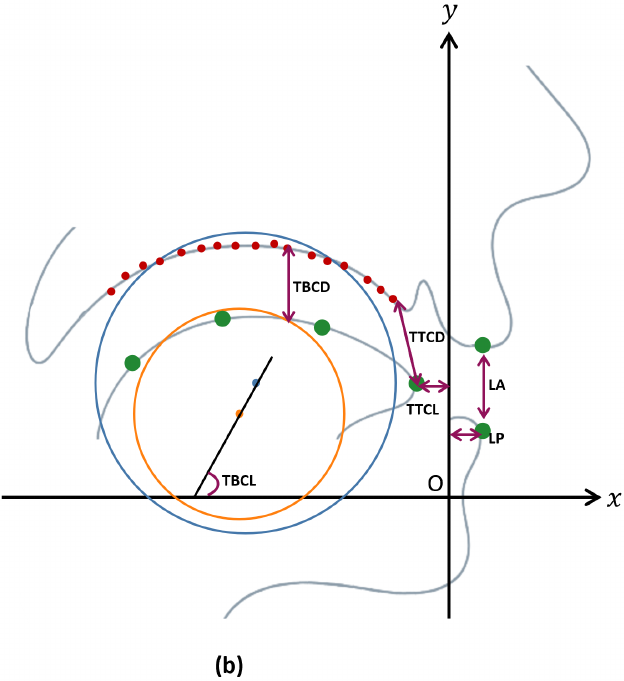}
    \caption{Illustration of the pellets' positions
    (left) and tract variables (right).} 
    \label{fig:pellets}
\end{figure*}
The markers above correspond to the global aspects of speech, where such characteristics are averaged over an entire, possibly long recording. A more recent approach considered the idea of acoustic \emph{landmarks}, i.e. ``abrupt articulatory events''~\cite{Huang2019} that take place at specific points in time. This makes it possible to investigate the effectiveness of biomarkers such as landmark counts and landmark $N$-grams, shown to identify depressed speakers with F1 Scores between 30\% and 40\% in the case of DAIC-WOZ, one of the most commonly used benchmarks in the literature~\cite{Huang2019b, Huang2019c}. In a similar vein, the approach by~\citet{Stasak2019} considers speech segments corresponding to different configurations of the articulators, especially vowels, and then extracts from them features known to represent speech signals effectively, e.g., the eGeMAPS~\cite{eyben2015}. The features are then fed to a classifier trained to detect depressed speakers and, when the performance is good, it means that the articulator configuration is a biomarker, i.e., that depressed and control speakers show different articulator configurations in correspondence of a particular sound. The results show that the best results are observed when considering speech segments corresponding to stress in intonation. The experiments by~\citet{Seneviratne2020} make a more explicit use of articulation by extracting TVs from the speech signals and by analyzing their correlation matrices with different delays. The results showed that the spectrum of the eigenvalues, the ordered set of eigenvalues obtained from the correlation matrix of the TVs, is different between depressed and control speakers (the distinction is based on a self-assessment questionnaire). 

A reliable biomarker is expected to provide information not only about the presence of a pathology, but also about its progression over time~\cite{Majkic2011}. However, the only work considering the latter over an extended period of time (18 months) and for a large number of speakers (more than 500) was presented by~\citet{Cummins2023}. The experiments considered 28 speech features and the results showed that the strongest markers are those related to speech rate, articulation rate and speech intensity in the case of a scripted speech task. The same problem is addressed by~\citet{Williamson2019} who focus on tracking the severity of depression over time. The analysis relied on the eigenvalues extracted from correlation matrices of the formants that were shown to be associated to the severity of the pathology. A similar approach was adopted by~\citet{Alpert2001}, where the focus was understanding the interplay between psychomotor retardation and different pharmacological therapies over time. Overall, these results confirm that the more severe the depression is, the slower a patient tends to speak.

\section{Extraction of the Biomarkers}\label{biomext}
In this section, we describe the process of extracting depression markers from speech signals in detail. Figure~\ref{fig:protocol} gives an overview of our approach.
\subsection{Speech Inversion}\label{speechinv}
The goal of the Speech Inversion step is to map every speech sample into a sequence of TV values extracted at regular time steps (see Section~\ref{tvdefinition} for the definition of the TVs). 
%
%
The inversion method used in this work is closely inspired by the approach proposed by~\citet{Sivaraman2019} (see Figure~\ref{fig:protocol}). The first step is to detect speech activity segments with \emph{WhisperX}~\cite{Bain2023}. A sequence of feature vectors composed of the first 13 \emph{Mel Frequency Cepstral Coefficients} (MFCC) is extracted from each speech segment using a 20 ms analysis window with a 10 ms overlap. These form the \emph{frames} as described in ~\citet{Sivaraman2019}.
Before being further processed, the MFCCs are $z$-normalized in line with the suggestions by~\citet{Mitra2012}.

The frames above are input to a Neural Network with 5 hidden layers (512 neurons each) with hyperbolic tangent as an activation function\footnote{The first three layers have the same parameters as those of the publicly available implementation available in~\cite{Sivaraman2019}, where the hidden layers are only 3 and not 5.}. The output layer has six neurons corresponding to the six TVs. The network is trained over the X-Ray Micro-Beam dataset (XRMB) ~\cite{Westbury1994}. This corpus has 46 speakers (21 male and 25 female) whose speech was recorded while being scanned in an X-Ray machine~\cite{Westbury1994}. This made it possible to track the positions of 6 pellets attached to \emph{Upper Lip} (UL), \emph{Lower Lip} (LL), \emph{Tongue Tip} (T1), \emph{Tongue Blade} (T2), \emph{Tongue Dorsum} (T3) and \emph{Tongue Root} (T4). Thus, the corpus includes a synchronised speech signal and the position of the pellets (in terms of $(x,y)$ coordinates over time). The speech has a sampling rate of 22.05 kHz and the pellet positions are sampled at 145 Hz. Figure~\ref{fig:pellets} (a) shows the location of the pellets within the speech production apparatus, and Section~\ref{tvdefinition} explains how the position of the pellets allows the calculation of the TV values.

The network used the original \texttt{train:val:test} split used by~\citet{Sivaraman2019,Siriwardena2023}, where 36 speakers were used for training, 5 for validation and 5 speakers for testing\footnote{The test speakers are JW18, JW31, JW33, JW39, and JW61, according to the IDs provided in~\cite{Westbury1994}.}. This allows testing whether the inversion results of our network match the state-of-the-art. Table~\ref{SI_results} shows the speech inversion results in terms of correlation between actual and predicted values of the TVs. Overall, the results suggest that the approach of this work reproduces the state-of-the-art in speech inversion. 
\begin{table*}[t]
\caption{Pearson correlation results comparing our speech inversion MLP model with previous studies. All correlations are statistically significant ($p<0.01$), and the best result per TV is in \textbf{bold}. The results reported for this article (bottom row) are the average over 10 repetitions corresponding to different random initializations of the neural network (the standard deviation is always lower than 0.005).}
\centering
\small
\begin{tabular}{lccccccc}
\cline{2-8}
 & \hspace{0.4cm}LA\hspace{0.4cm} & \hspace{0.4cm}LP\hspace{0.4cm} & \hspace{0.2cm}TBCL\hspace{0.2cm} & \hspace{0.2cm}TBCD\hspace{0.2cm} & \hspace{0.2cm}TTCL\hspace{0.2cm} & \hspace{0.2cm}TTCD\hspace{0.2cm} & \hspace{0.2cm}Average\hspace{0.2cm} \\ 
\toprule
\citealt{Sivaraman2019} & 0.856 & 0.613 & \textbf{0.866} & 0.745 & 0.707 & 0.907 & 0.782 \\
\citealt{attia2023enhancing} & 0.868 & 0.590 & 0.742 & 0.780 & 0.597 & 0.893 & 0.745 \\

\citealt{2024improving} w/ MFCCs & 0.860 & 0.710 & 0.742 & 0.775 & 0.742 & 0.898 & 0.788 \\

\citealt{2024improving} w/ HuBERT\hspace{0.4cm} & \textbf{0.878} & \textbf{0.724} & 0.743 & 0.809 & \textbf{0.787 }& \textbf{0.925} & 0.810 \\

This article & 0.855 & 0.714 & 0.816 &\textbf{0.841} & 0.773 & 0.897 & \textbf{0.816} \\ 
\bottomrule
\end{tabular}
\label{SI_results}
\end{table*}

\subsection{Definition of the TVs}\label{tvdefinition}
Figure~\ref{fig:pellets} shows the position of the 6 pellets (\emph{Upper Lip} (UL), \emph{Lower Lip} (LL), \emph{Tongue Tip} (T1), \emph{Tongue Blade} (T2), \emph{Tongue Dorsum} (T3), \emph{Tongue Root} (T4)) and the TV measurements. Figure~\ref{fig:pellets}  (a) also shows the approximations commonly used to represent the tongue and the palate, namely the Tongue Circle and the Palate Circle. The Tongue Circle is estimated as a circle fitted to the pellets T2, T3, and T4 with a fixed radius of 20 mm following the protocol of previous work~\cite{Sivaraman2019}, while the Palate Circle is obtained by fitting a circle to the palate points available for each speaker using a least-squares optimization procedure. Let the centre of the Tongue Circle be denoted as $\vec t = (x_T,y_T)$ and that of the Palate Circle be $\vec p = (x_P,y_P)$. The TVs, illustrated in Figure~\ref{fig:pellets} (b), are functions of the pellets' coordinates and are defined as follows:
\begin{itemize}
\item Lip Aperture (LA) is the Euclidean distance between the positions of the LL and UL pellets, denoted with $\vec u_{LL}=(x_{LL},y_{LL})$ and $\vec u_{UL}=(x_{UL},y_{UL})$, respectively.
\begin{equation}\text{LA}=\Vert\vec u_{LL}-\vec u_{UL}\Vert_2;\end{equation} 
\item Lip Protrusion (LP) is the horizontal distance of the LL pellet from the origin along the $x$ axis (see Fig~\ref{fig:pellets}):
\begin{equation}
\text{LP} = x_{LL};
\end{equation}
\item Tongue Body Constriction Location (TBCL) is measured as the angle between the $x$ axis and the line passing through $\vec t$ and $\vec p$:
\begin{equation}
\text{TBCL} = \arctan \left( \frac{y_{T} - y_{P}}{x_{T} - x_{P}} \right);
\label{eq:TBCL}
\end{equation}
\item Tongue Body Constriction Degree (TBCD) is the shortest distance between any point $\vec u_T$ on the Tongue Circle between $\vec u_{T2}$ and $\vec u_{T4}$ (see Figure~\ref{fig:pellets}) and any point $\vec u_P$ on the Palate Circle:
\begin{equation}
\text{TBCD} = \min_{\vec u_P,\vec u_T} \Vert\vec u_P-\vec u_T\Vert_2;   
\end{equation} 
\item Tongue Tip Constriction Location (TTCL) is the horizontal distance of the T1 pellet position, denoted with $\vec u_{T1} = (x_{T1},y_{T1})$ from the $y$ axis:
\begin{equation}
\text{TTCL} = x_{T1};
\label{eq:TTCL}
\end{equation}
\item Tongue Tip Constriction Degree (TTCD) is the minimum distance between $\vec u_{T1}$ and a point $\vec u_P$ of the Palate Circle:
\begin{equation}
\text{TTCD} = \min_{\vec u_P} \Vert\vec u_{T1}-\vec u_P\Vert_2.
\label{eq:TTCD}
\end{equation}
\end{itemize}
The speech inversion process described above converts the speech signal into a sequence of these six TVs that measure the locations of the articulatory parameters.
\subsection{Embedding Extraction from the TVs}\label{nonlinear}
\begin{figure*}[t]
    \centering
    \includegraphics[width=0.9\textwidth]{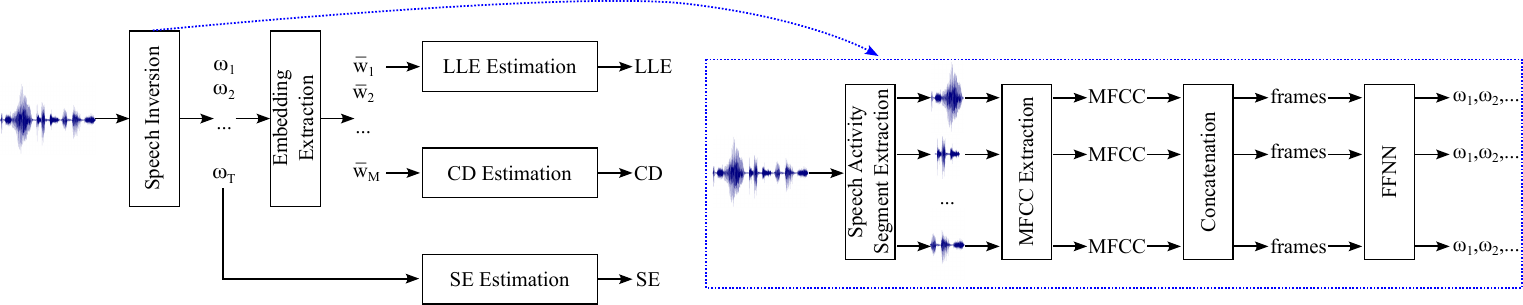}
    \caption {The figure shows our experimental approach. Every recording is mapped into a sequence $\omega_1, \ldots, \omega_T$ of TV values extracted at regular time steps through a Speech Inversion step detailed in the right part of the picture (see Section~\ref{speechinv}). The sequence is used to obtain the state-space embeddings $\vec w_1,\ldots,\vec w_M$ that allow one to estimate LLE and CD. The sequence is also used to estimate an information-theoretic metric SE.}
    \label{fig:protocol}
\end{figure*}
The TV values are extracted at regular time steps, where each of the six TVs yields a sequence $\{\omega_k\}_{k=1}^T$, where $\omega_k \in \mathbb{R}$ is the TV value extracted at time step $k$ with $T$ being the total number of steps. The Takens' Theorem~\cite{Takens2006} suggests that the sequence \{$\omega_k$\} contains information about the state-space of the dynamical system underlying the TVs, i.e., the articulation process. Following this theorem, the state-space can be reconstructed through the \textbf{embeddings} (different from the commonly understood concept of embeddings in Machine Learning)~\cite{Tan2023}. The embeddings are essentially subsequences derived from the original sequence: $\vec w_k=\{\omega_k,\omega_{k+\tau},\ldots,\omega_{k+(d-1)\tau}\}$, where $\tau$ and $d$ are the \emph{delay} and \emph{dimension} of the embeddings. The embedding extraction step (see Figure~\ref{fig:protocol}) thus maps every sequence of TV values into a set of embeddings $\mathcal{W}=\{\vec w_1,\ldots,\vec w_M\}$, where $M$ is the total number of embeddings available for a given TV sequence depending on the parameters $\tau$ and $d$. The sequence $\mathcal{W}$ is the trajectory of the dynamical system in the state-space, a representation of the way the system evolves over time.  

In the experiments of this work, the parameter $\tau$ was selected by minimizing the mutual information between $\omega_k$ and $\omega_{k+\tau}$ with the publicly available package \emph{Neurokit2}~\cite{Makowski2021}\footnote{\url{https://neuropsychology.github.io/NeuroKit/functions/complexity.html\#neurokit2.complexity.complexity_delay}, last accessed in September 2025.}. This ensures that the delay is long enough to limit the redundancy between the information conveyed by the different components of an embedding~\cite{Fraser1986}. The other parameter $d$ was determined using the approach proposed by~\citet{Cao1997} that examines how the average distance between the neighbouring points in the reconstructed state-space changes when the dimension increases from $d$ to $(d + 1)$. The optimal dimension is identified when increasing $d$ does not change significantly the average distance indicating that no additional information is created as we move to a higher dimension. The approach was implemented with the \emph{Neurokit2} package~\cite{Makowski2021}\footnote{\url{https://neuropsychology.github.io/NeuroKit/functions/complexity.html\#neurokit2.complexity.complexity_dimension}, last accessed in September 2025.}.
\subsection{Largest Lyapunov Exponent Estimation}
\label{sec:lle}
Consider the sequence $\mathcal{W}=\{\vec w_1,\vec w_2,\ldots,\vec w_M\}$, where $M$ denotes the number of embeddings. This trajectory represents the temporal evolution of the underlying dynamical system, i.e., the articulatory process. In principle, the same system should follow roughly the same trajectory when starting from the same point of the state-space. However, in the presence of chaos, trajectories having the same starting point can look very different, even if the underlying system is the same. Lyapunov Exponents ``quantify the exponential divergence of initially close state-space trajectories and estimate the amount of chaos in a system''~\cite{Rosenstein1993}. Therefore, this work involves measuring the \emph{Largest Lyapunov Exponent} (LLE) of $\mathcal{W}$ that is known to characterize the rate of exponential divergence~\cite{Eckmann1985}.

This work is based on the LLE estimation approach proposed by~\citet{Rosenstein1993} using the publicly available package \emph{Neurokit2}~\cite{Makowski2021}\footnote{\url{https://neuropsychology.github.io/NeuroKit/functions/complexity.html\#complexity-lyapunov}, last accessed in September 2025.}. For each embedding $\vec w_j$, its nearest neighbour $\mathcal{N}(\vec w_j)$ is identified subject to a temporal separation constraint $\Delta t\geq \frac{1}{F}$, where $F$ is the sampling frequency. The initial separation is defined as $d_j(0) = \|\vec {w}_j - \mathcal{N}(\vec{w}_j)\|_2$. The evolution of the distance between neighbouring embeddings after $i$ discrete time steps is then denoted as
$d_j(i) = \|\vec{w}_{j+i} - \mathcal{N}(\vec{w}_j) _{+ i}\|_2$. Under the assumption of exponential divergence, these distances evolve approximately as $d_j(i) \approx d_j(0)\, e^{\lambda_1 i \Delta t}$, which gives $\log d_j(i) \approx \log d_j(0) + \lambda_1 i \Delta t$. Thus, for each reference point $j$, the logarithmic divergence grows linearly with time with slope $\lambda_1$. The estimate of Largest Lyapunov Exponent (LLE) for finite time steps $i$ (scale) is given by averaging over j with time step $i$:
\begin{equation}
\lambda_1(i) = \frac{1}{i\Delta t} \cdot \frac{1}{M-i} \sum_{j=1}^{M-i} \log \frac{d_j(i)}{d_j(0)}
\label{eq:lle}
\end{equation}
The final LLE, $\lambda_1$, is obtained as the slope of the best linear fit of the mean logarithmic divergence curve:
$\left\langle \log d(i) \right\rangle = \frac{1}{M-i}\sum_{j=1}^{M-i} \log d_j(i)$
with respect to time $i \Delta t$. Hence, the final estimate can be written as:
\begin{equation}
\lambda_1 = \frac{d}{dt} \left\langle \log d(i) \right\rangle
\end{equation}
%
\subsection{Correlation Dimension Estimation}
The \emph{Correlation Dimension} (CD) measures the minimum dimensionality needed to represent a trajectory $\mathcal{W}$ in the state-space. Higher CD means that the trajectory is less constrained and the system has higher degrees of freedom. In the case of the TVs, this means greater variability and flexibility in the articulation process. This work estimated the CD using the method proposed by~\citet{Grassberger1983}, which relies on computing the correlation sum $C(\rho)$. This is defined as the fraction of the available embedding pairs $(\vec w_i,\vec w_j)$ that are separated by a distance lower than a threshold $\rho$: 
\begin{equation}
C(\rho) = \frac{2}{M(M - 1)} \sum_{i=1}^{M} \sum_{j=i+1}^{M} \Theta\left(\rho - \Vert\vec w_i-\vec w_j\Vert_2 \right),
\label{eq:correlation_integral}
\end{equation}
where $M$ is the total number of embeddings and $\Theta$ is the Heaviside function (equals $1$ when its argument is positive and $0$ otherwise). The quantity $C(\rho)$ is related to the CD, denoted by $D$, as $C(\rho)\propto \rho^D$, meaning that $\ln C(\rho) \propto D\ln \rho$. $C(\rho)$ is computed for different values of $\rho$, and the value of $D$ is defined as the slope of the linear relationship between $\ln C(\rho)$ and $\ln \rho$. This slope can be computed by using any line fitting method.
\subsection{Sample Entropy Estimation}
The \emph{Sample Entropy} (SE) aims at estimating ``the randomness of a series of data without any previous knowledge about the source generating the dataset''~\cite{Delgado-Bonal2019} and it ``also indicates more self-similarity in the time series''~\cite{Richman2000}, i.e., how fast new patterns emerge. Overall, low SE indicates that the TV sequences tend to be repetitive and always show similar patterns. In contrast, when the SE is high, new information is expected to generate at a higher rate, where new patterns emerge. SE has been shown to be a useful marker of facial motion-related atypicality in Autism \cite{Guha2018}, but has not been investigated in the context of speech to the best of our knowledge.

The estimation of the SE~\cite{Richman2004}  takes place over a temporal sequence $\{\omega_k\}_{k=1}^K$ of TV values extracted from a speech recording, where $K$ is the length of the sequence, $\omega_k$ is the TV value extracted at time $k/F$ and $F$ is the TV sampling frequency. The first step of the process is to extract $(K-m+1)$ vectors $\vec \omega_m(k) = (\omega_k,\omega_{k+1},\ldots,\omega_{k+m-1})$ from the sequence, where $m$ is a parameter to be set. The second step is to count the number $B$ of pairs $[\vec \omega_m(i),\vec \omega_m(j)]$ such that $dist(\vec \omega_m(i),\vec \omega_m(j))\leq r$, where $r$ is a user defined threshold, $dist(.)$ is a distance function. In this work, we define $dist(\vec \omega_m(i),\vec \omega_m(j))=\max_{k\in[1,m]} |\omega_m(i,k)-\omega_m(j,k)|$, and $\omega_m(i,k)$ is the $k^{th}$ component of vector $\vec \omega_m(i)$; any other distance function is also valid. The third step of the process is to count the number $A$ of pairs $[\vec \omega_{m+1}(i),\vec \omega_{m+1}(j)]$ such that $dist(\vec \omega_{m+1}(i),\vec \omega_{m+1}(j))\leq r$. The resulting SE is given by:
\begin{equation}
SE(m,r,K) = -\ln\left( \frac{A}{B}\right ),
\end{equation}
where the parameter $m$ corresponds to the dimension of the state-space embeddings (estimated with the method described earlier) and $r$ is set to one-fifth of the standard deviation of the sequence of the TV values. 
\section{Experiments and Results}\label{expres}
The goal of this work is to identify depression biomarkers in speech, i.e., measurable aspects of speech that account for the presence of the pathology. The experiments thus involve comparison between healthy controls and speakers diagnosed with depression in terms of LLE, CD and SE to identify statistically significant differences between the two groups. The rest of this section presents the dataset used in the experiments and the results that were obtained.

\subsection{The Androids Corpus}
All experiments were performed over the Androids Corpus~\cite{Tao2023}, a publicly available collection of 228 speech recordings produced by 118 speakers, including 64 diagnosed with depression by professional psychiatrists\footnote{\url{https://github.com/androidscorpus/data}, last accessed in September 2025.\label{androidsfootnote}}. The latter aspect is of particular importance because it means that the results presented in this work depend on the actual presence (or absence) of depression and not, like it happens in most previous works, on scores resulting from self-assessment questionnaires~\cite{Cummins2023}. In addition, depressed and control speakers of the Corpus are matched in terms of age, gender and education level distribution (see Table~\ref{demographics}). This reduces the potential influence of demographic confounding factors on articulatory dynamics and allows uncovering differences associated with the mental health condition rather than demographic variability.

The speakers performed two different tasks, referred to as \emph{Reading Task} and \emph{Interview Task}. The Reading Task corresponds to reading a text (\emph{The North Wind and the Sun} by Aesop), while the second corresponds to answering questions about everyday life (the same for all speakers and in the same order). The first task aims at collecting read speech, while the other one allows the collection of spontaneous speech. The key difference between the two is that the cognitive load required to plan what to say next is limited, if not absent, when reading and significant when speaking spontaneously. This is important because the cognitive load can blur the difference between depressed and control speakers~\cite{Alpert2001}, especially when it comes to speaking rate and length or frequency of pauses, some of the biomarkers that were investigated the most in literature (see Section~\ref{survey}).   

In case of the Reading Task, the total duration of the recordings is 1 hour, 33 minutes and 49 seconds. The average recording duration for a subject is 50.3 s 
overall, for the speakers with depression, this average is 52.9 s and for the controls the average is 47.4 s. In case of Interviews, the total duration of audio recording is 7 hours, 24 minutes and 22 seconds with an average length of 229.8 s per session. For the speakers with depression, the average is 198.8 s, while it is 268.0 s for the control group.
\begin{table}[t!]
\centering
\caption {Demographic information about the Androids Corpus. Mean age $\pm$ standard deviation for each group is shown per task. Acronyms \emph{F} and \emph{M} stand for Female and Male, respectively. Acronyms \emph{L} and \emph{H} stand for Low (8 years of study at most) and High (at least 13 years of study) education level, respectively.
The sum over the education level columns does not correspond to the total number of participants (118) because 2 of these did not provide details about their studies. }
\label{demographics}
 \begin{tabular}{l l ccccc}
 \hline
 Task & & Age & M & F & L & H \\ 
 \hline
 \multirow{3}*{Reading}
 
& Control & 47.1 $\pm$ 12.8 & 12 & 42 & 19 & 35 \\
& Depression & 47.4 $\pm$ 11.9 & 20 & 38 & 25 & 32 \\
& Total & 47.2 $\pm$ 12.3 & 32 & 80 & 44 & 67 \\
\hline
\multirow{3}*{Interview}
& Control & 47.3 $\pm$ 12.7 & 11 & 41 & 19 & 33 \\
& Depression & 47.5 $\pm$ 11.6 & 21 & 43 & 29 & 33 \\
& Total & 47.4 $\pm$ 12.1 & 32 & 84 & 48 & 66 \\
 \hline
\multirow{3}*{Total}
& Control & 47.3 $\pm$ 12.7 & 11 & 41 & 19 & 33 \\
& Depression & 47.4 $\pm$ 11.9 & 20 & 38 & 25  & 32 \\
& Overall & 47.3 $\pm$ 12.2 & 31 & 79 & 44 & 65 \\
\hline
\end{tabular}
\end{table}

\subsection{Results}
Table~\ref{lleresults} shows the average ($\pm$ standard deviation) LLE for depressed and control speakers. The results are reported separately for the two types of speech recordings (read and spontaneous) corresponding to the Reading and Interview tasks. The comparison between the depressed and control speakers is performed for each TV individually, for all TVs together, and for two subsets corresponding to lips (LA and LP) and tongue region (TBCL, TBCD, TTCL and TTCD). Overall, Table~\ref{lleresults} suggests that, in the case of read speech, the differences in LLE values between the two groups are statistically significant in all cases except LA. Besides reporting the p-values from the Mann-Whitney test, the table provides an assessment of the effect size through the absolute value of the Cliff's $\delta$, given by:
\begin{equation}
\delta = \frac{n(x_i>y_j)-n(x_i<y_j)}{N_xN_y},    
\end{equation}
%
where $N_x$ is the number of values in a set $\mathcal{X}=\{x_1,\ldots,x_{N_x}\}$, $N_y$ is the number of values in a set $\mathcal{Y}=\{y_1,\ldots,y_{N_y}\}$, $n(x_i>y_j)$ is the number of pairs $(x_i,y_j)$ in which $x_i>y_j$, and $n(x_i<y_j)$ is the number of pairs $(x_i,y_j)$ in which $x_i<y_j$. Cliff’s $\delta$ can be interpreted as the degree to which quantities in one group tend to be larger than those in the other group. For example, $\delta=0$ indicates that values from either group is equally likely to be larger than the other, which corresponds to no effect. In contrast, a value such as $\delta=0.5$ indicates that 75\% of the paired comparisons favour one group over the other. 

According to the terminology proposed by~\citet{Meissel2024}, the $\delta$ value is at least \emph{Medium} in 6 cases out of the total 9. In particular, according to the same terminology, the effect is \emph{Large} when using all TVs or the subset corresponding to the Tongue. This seems to suggest that, while working well individually, the TVs are effective biomarkers especially when considered jointly. One possible explanation is that the TVs account for different aspects of the same articulation process and they all jointly contribute to shaping speech. As a confirmation, in the case of spontaneous speech, statistically significant differences are observed only for TV sets (Lips and Tongue) or all TVs together, although with a smaller Cliff's $\delta$. 

The LLE can be thought of as a ``measure of a system’s predictability''~\cite{Rudisuli2013}: the higher the LLE, the less a system is predictable. Table~\ref{lleresults} shows that, whenever there is a statistically significant effect, the LLE is greater for depressed speakers. One possible reason is that speech production is the most complex motor process in humans and many people affected by depression ``exhibit motor programming disturbances''~\cite{Caligiuri2000}. These probably make the articulation process less predictable and, as a confirmation of such hypothesis, previous work showed that, in depressed speakers, the values of the same feature extracted at different times tend to correlate less than in control speakers~\cite{Quatieri2012,Williamson2013}, possibly due to lack of articulatory coordination~\cite{Williamson2019}. 

\begin{table*}[t!]
\caption{The table reports, for every TV or group of TVs, average and standard deviation of the LLE observed for Control and Depressed speakers, for both Reading and Interview Task. The \emph{p-value} column reports the outcome of the Mann-Whitney test (the values in bold are statistically significant after the application of the False Discovery Rate correction proposed by Benjamini and Hochberg~\cite{benjamini1995}). Column $|\delta|$ shows the absolute value of the Cliff's $\delta$. Group \emph{Lips} includes LA and LP, while group \emph{Tongue} includes  TBCL, TBCD, TTCL and TTCD.}\label{lleresults}
\setlength{\tabcolsep}{4pt}
\begin{tabular}{l|cccc||cccc}
\multicolumn{1}{c}{}
 & \multicolumn{4}{c}{Reading Task} 
 & \multicolumn{4}{c}{Interview Task}\\ 
\toprule
TV & Control & Depressed & p-value & $|\delta|$ & Control & Depressed & p-value & $|\delta|$ \\\hline
LA     & 0.021$\pm$0.005 & 0.023$\pm$0.006 & 0.104 & 0.18 & 0.029$\pm$0.005 & 0.032$\pm$0.007 & 0.023 & 0.23 \\
LP     & 0.020$\pm$0.005 & 0.025$\pm$0.007 & \textbf{$<$0.001} & 0.38 & 0.032$\pm$0.006 & 0.033$\pm$0.006 & 0.234 & 0.09 \\
TTCL   & 0.022$\pm$0.006 & 0.026$\pm$0.007 & \textbf{0.001} & 0.34 & 0.028$\pm$0.005 & 0.030$\pm$0.006 & 0.035 & 0.22 \\
TTCD   & 0.020$\pm$0.004 & 0.024$\pm$0.006 & \textbf{0.002} & 0.28 & 0.028$\pm$0.005 & 0.032$\pm$0.008 & 0.013 & 0.26 \\
TBCL   & 0.022$\pm$0.006 & 0.029$\pm$0.008 & \textbf{$<$0.001} & 0.41 & 0.033$\pm$0.007 & 0.035$\pm$0.008 & 0.118 & 0.15 \\
TBCD   & 0.025$\pm$0.008 & 0.030$\pm$0.009 & \textbf{$<$0.001} & 0.37 & 0.034$\pm$0.005 & 0.036$\pm$0.007 & 0.045 & 0.19 \\ \hline
Lips   & 0.021$\pm$0.004 & 0.024$\pm$0.005 & \textbf{0.003} & 0.31 & 0.030$\pm$0.006 & 0.032$\pm$0.006 & \textbf{0.049} & 0.19 \\
Tongue & 0.022$\pm$0.004 & 0.027$\pm$0.006 & \textbf{$<$0.001} & 0.51 & 0.031$\pm$0.005 & 0.033$\pm$0.006 & \textbf{0.030} & 0.23 \\
All & 0.022$\pm$0.003 & 0.026$\pm$0.005 & \textbf{$<$0.001} & 0.50 & 0.031$\pm$0.005 & 0.033$\pm$0.006 & \textbf{0.024} & 0.23 \\
\bottomrule
\end{tabular}
\end{table*}

\begin{table*}[t]
\caption{The table reports, for every TV or group of TVs, average and standard deviation of the CD observed for Control and Depressed speakers, for both Reading and Interview Task. The \emph{p-value} column reports the outcome of the Mann-Whitney test. The values in bold are statistically significant after the application of the False Discovery Rate correction proposed by Benjamini and Hochberg~\cite{benjamini1995}. Column $|\delta|$ shows the absolute value of the Cliff's $\delta$. Group \emph{Lips} includes LA and LP, while group \emph{Tongue} includes  TBCL, TBCD, TTCL and TTCD.}\label{cdresults}

\setlength{\tabcolsep}{6pt}
\begin{tabular}{l|cccc||cccc}
\multicolumn{1}{c}{}
 & \multicolumn{4}{c}{Reading Task} 
 & \multicolumn{4}{c}{Interview Task}\\ 
\toprule
TV & Control & Depressed & p-value & $|\delta|$ & Control & Depressed & p-value & $|\delta|$ \\\hline
LA     & 3.66$\pm$0.32 & 3.47$\pm$0.31 & \textbf{0.001} & 0.33 & 3.19$\pm$0.23 & 3.10$\pm$0.28 & \textbf{0.006} & 0.27 \\
LP     & 3.62$\pm$0.29 & 3.47$\pm$0.27 & \textbf{0.008} & 0.26 & 3.30$\pm$0.21 & 3.21$\pm$0.31 & \textbf{0.020} & 0.22 \\
TTCL   & 3.42$\pm$0.33 & 3.34$\pm$0.37 & \textbf{0.022} & 0.22 & 3.20$\pm$0.24 & 3.06$\pm$0.36 & \textbf{0.013} & 0.24 \\
TTCD   & 3.35$\pm$0.36 & 3.16$\pm$0.36 & \textbf{0.004} & 0.29 & 2.90$\pm$0.21 & 2.73$\pm$0.30 & \textbf{$<$0.001} & 0.36 \\
TBCL   & 3.16$\pm$0.33 & 2.94$\pm$0.31 & \textbf{$<$0.001} & 0.37 & 2.63$\pm$0.18 & 2.50$\pm$0.28 & \textbf{0.002} & 0.32 \\
TBCD   & 3.45$\pm$0.36 & 3.24$\pm$0.38 & \textbf{0.001} & 0.34 & 2.89$\pm$0.21 & 2.74$\pm$0.27 & \textbf{0.001} & 0.33 \\ \hline
Lips   & 3.64$\pm$0.22 & 3.47$\pm$0.23 & \textbf{$<$0.001} & 0.40 & 3.25$\pm$0.19 & 3.14$\pm$0.25 & \textbf{0.007} & 0.27 \\
Tongue & 3.35$\pm$0.23 & 3.17$\pm$0.25 & \textbf{$<$0.001} & 0.40 & 2.91$\pm$0.15 & 2.77$\pm$0.25 & \textbf{$<$0.001} & 0.36 \\[3pt]
Overall& 3.46$\pm$0.17 & 3.28$\pm$0.21 & \textbf{$<$0.001} & 0.49 & 3.04$\pm$0.14 & 2.91$\pm$0.21 & \textbf{0.001} & 0.35 \\
\bottomrule
\end{tabular}
\end{table*}

Table~\ref{cdresults} shows the CD values for depressed and control speakers in correspondence of individual TVs as well as groups of TVs corresponding to tongue and lips or all TVs together. The CD is an upper bound of the dimensionality needed to represent the points that belong to a set. In the experiments of this work, the points represent the states of a dynamical system and, therefore, the CD is the dimensionality of an \emph{attractor}~\cite{Grassberger1983}, the set of the states that a dynamical system tends to reach. The lower the dimensionality, the less the degrees of freedom of a system. In other words, when the dimension of an attractor is low, the underlying dynamical system shows a lower degree of variability in its observable behaviour. 

Table~\ref{cdresults} shows that, in the case of control speakers, the CD is always greater to a statistically significant extent, whether the comparison is made for individual TVs, for TV groups or for all TVs. Furthermore, the table shows that, unlike the LLE, the CD appears to be an effective marker for both read and spontaneous speech. This suggests that control speakers display, on average, higher variability in their speech patterns. One possible explanation is that the literature shows that, on average, depressed speakers manifest lower variability along multiple aspects of speech, including energy~\cite{Quatieri2012,Cummins2013}, spectral features~\cite{Cummins2013b}, prosody (pitch, energy and speaking rate)~\cite{Moore2003}, frequency range in vowel production~\cite{Scherer2016}, etc. Given that articulation is the process that underlies all of these observable aspects of speech, the CD captures lower variability for all of them jointly. This probably explains why statistically significant differences are observed for all TVs.
\begin{table*}[t]
\caption{The table reports, for every TV or group of TVs, average and standard deviation of the SE observed for Control and Depressed speakers, for both Reading and Interview Task. The \emph{p-value} column reports the outcome of the Mann-Whitney test. The values in bold are statistically significant after the application of the False Discovery Rate correction proposed by Benjamini and Hochberg~\cite{benjamini1995}. Column $|\delta|$ shows the absolute value of the Cliff's $\delta$. Group \emph{Lips} includes LA and LP, while group \emph{Tongue} includes  TBCL, TBCD, TTCL and TTCD.}\label{seresults}

\setlength{\tabcolsep}{6pt}
\begin{tabular}{l|cccc||cccc}
\multicolumn{1}{c}{}
 & \multicolumn{4}{c}{Reading Task} 
 & \multicolumn{4}{c}{Interview Task}\\ 
\toprule
TV & Control & Depressed & p-value & $|\delta|$ & Control & Depressed & p-value & $|\delta|$ \\ 
\toprule
LA     & 0.26$\pm$0.06 & 0.23$\pm$0.05 & \textbf{0.001} & 0.33 & 0.22$\pm$0.03 & 0.20$\pm$0.03 & \textbf{0.001} & 0.33 \\
LP     & 0.22$\pm$0.04 & 0.19$\pm$0.04 & \textbf{$<$0.001} & 0.39 & 0.19$\pm$0.03 & 0.17$\pm$0.03 & \textbf{0.002} & 0.32 \\
TTCL   & 0.24$\pm$0.06 & 0.23$\pm$0.05 & 0.202 & 0.09 & 0.24$\pm$0.03 & 0.23$\pm$0.04 & \textbf{0.024} & 0.21 \\
TTCD   & 0.26$\pm$0.05 & 0.25$\pm$0.06 & 0.119 & 0.13 & 0.24$\pm$0.03 & 0.23$\pm$0.04 & 0.213 & 0.09 \\
TBCL   & 0.18$\pm$0.04 & 0.18$\pm$0.05 & 0.487 & 0.00 & 0.13$\pm$0.02 & 0.12$\pm$0.04 & 0.087 & 0.15 \\
TBCD   & 0.16$\pm$0.04 & 0.14$\pm$0.04 & \textbf{$<$0.001} & 0.30 & 0.13$\pm$0.02 & 0.11$\pm$0.02 & \textbf{0.001} & 0.33 \\ \hline
Lips   & 0.24$\pm$0.04 & 0.21$\pm$0.05 & \textbf{0.002} & 0.42 & 0.20$\pm$0.03 & 0.18$\pm$0.03 & \textbf{$<$0.001} & 0.38 \\
Tongue & 0.21$\pm$0.03 & 0.20$\pm$0.03 & \textbf{0.024} & 0.22 & 0.20$\pm$0.02 & 0.19$\pm$0.03 & \textbf{0.022} & 0.22 \\[3pt]
Overall& 0.22$\pm$0.03 & 0.21$\pm$0.03 & \textbf{$<$0.001} & 0.37 & 0.20$\pm$0.02 & 0.19$\pm$0.02 & \textbf{0.001} & 0.33 \\
\bottomrule
\end{tabular}
\end{table*}

\begin{figure*}[t]
    \centering
    \includegraphics[width=0.32\textwidth]{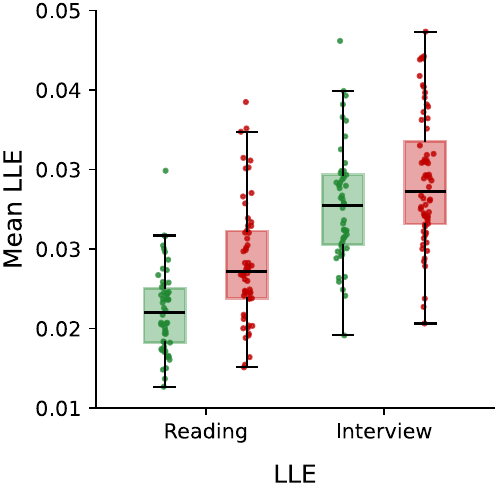}
    \includegraphics[width=0.32\textwidth]{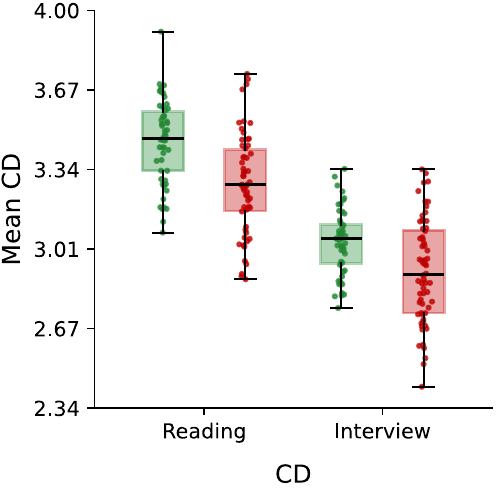}
    \includegraphics[width=0.32\textwidth]{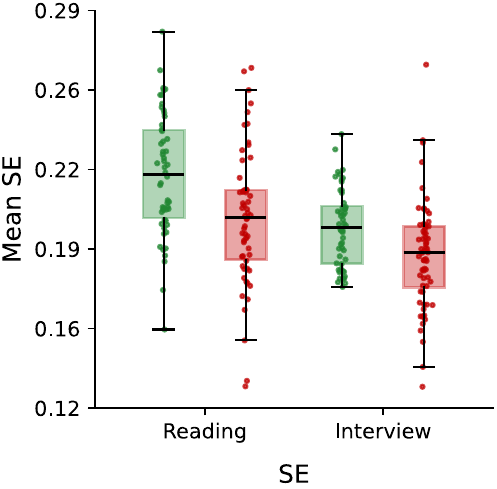}
    \caption{Distribution of speaker-level mean LLE, CD and SE for the Reading and Interview tasks. 
    For each task, boxplots illustrate the distribution across speakers for healthy controls (green) and depressed patients (red). Individual points represent speaker-level mean values averaged across the six tract variables.}
    \label{fig:boxplots}
\end{figure*}

Table~\ref{seresults} shows the results corresponding to the SE. The difference between depressed and control speakers is statistically significant for all TV groups and for all individual TVs except TTCL, TTCD and TBCL for the Reading Task and except TTCD and TBCL for the Interview Task. The SE accounts for ``the randomness of a series of data''~\cite{Delgado-Bonal2019} and the table shows that its values are consistently greater for the control speakers. This suggests that depressed speakers show a lower degree of randomness, an observation that, at first sight, seems to contradict the results obtained with the LLE (see above). However, the two markers capture different aspects because the LLE measures how predictable the speech patterns are, while the SE measures how repetitive they are, i.e., how much the same patterns tend to reappear in a series (the speech signal in this case). In this respect, the results presented in Table~\ref{lleresults} and Table~\ref{seresults} are compatible with each other: LLE shows statistically significant effects mostly for read speech, and the SE shows significant effects for both read and spontaneous speech. Figure \ref{fig:boxplots} summarizes the main trends observed in Tables \ref{lleresults}-\ref{seresults}. The distributions show that depressed speakers generally have higher LLE values, indicating lower predictability, while control speakers tend to have higher CD and SE values, indicating greater variability and less repetitive articulatory patterns.

\subsection{Gender Differences}
Table~\ref{fmcomparison} shows the absolute values of the Cliff's $\delta$ whenever there is a statistically significant difference between depressed speakers and the controls (the symbol $\times$ means that the difference is \emph{not} statistically significant). This investigates whether or not the observed effect is influenced by the speaker's (reported) gender. 
The results suggest that the proposed markers tend to be more effective in the case of female speakers. In fact, the cases of statistically significant differences for female speakers are over four times those of the male speakers (out of a total of 54 cases, 10 for male and 45 for female). However, the probable reason is that the number of female speakers is greater (80 vs 32 for the Reading Task and 84 vs 32 for the Interview Task) due to the highest prevalence of depression among women~\cite{Kuehner2017}. As a confirmation, when performing the same comparisons using only 32 randomly selected female speakers, the number of statistically significant effects becomes comparable to the one observed for the male ones (3 out of 54).

Another interesting observation is that, when considering all speakers, the LLE shows 8 statistically significant differences for read speech and 3 for spontaneous speech (see Table~\ref{lleresults}), while for female speakers both numbers are 8. 
In other words, the LLE appears to be an effective marker for both read and spontaneous speech in the case of female speakers, but not in the case of male ones. One possible explanation is that the cognitive load required to plan what to say next - the possible explanation between the effectiveness differences observed between read and spontaneous speech~\cite{Alpert2001} - does not have the same impact on both female and male speakers. 

Note that for the male speakers in the Androids Corpus the difference between the controls and the depressed group is less pronounced. It is also observed that the $|\delta|$ values tend to be higher for male speakers; this trend holds across all 10 statistically significant effects reported in Table~\ref{fmcomparison}. In contrast, only 6 out of the 45 $|\delta|$ values computed for female speakers exceed the minimum $|\delta|$ value observed for males, which is 0.41. This suggests that, on average, the differences between depressed speakers and control participants are more pronounced in female speakers, at least with respect to the markers proposed in this study.

\begin{table*}[t!]
\caption{The table shows the difference between depressed and control groups (in terms of $|\delta|$ values) when considering female and male speakers separately. The acronym F stands for female, while the M one stands for male. The symbol $\times$ means that the difference between depressed and control speakers is not statistically significant according to a Mann-Whitney test with False Discovery Rate correction~\cite{benjamini1995}.}\label{fmcomparison}
\setlength{\tabcolsep}{3pt}
\begin{tabular}{l|cc|cc|cc||cc|cc|cc}
\multicolumn{1}{c}{}
 & \multicolumn{6}{c}{Reading Task} 
 & \multicolumn{6}{c}{Interview Task}\\ 
\toprule
TV & LLE(F) & LLE(M) & CD(F) & CD(M) & SE(F) & SE(M) & LLE(F) & LLE(M) & CD(F) & CD(M) & SE(F) & SE(M)\\
\hline
LA   & $\times$ & $\times$ & 0.23     & $\times$ & 0.26     & 0.50     & 0.28     & $\times$ & 0.30     & $\times$ & 0.30 & $\times$\\
LP   & 0.44     & $\times$ & 0.32     & $\times$ & 0.30     & 0.60     & $\times$ & $\times$ & $\times$ & $\times$ & 0.37 & $\times$\\
TTCL & 0.26     & $\times$ & $\times$ & $\times$ & $\times$ & $\times$ & 0.34     & $\times$ & $\times$ & $\times$ & 0.25 & $\times$ \\
TTCD & 0.25     & $\times$ & 0.27     & $\times$ & $\times$ & $\times$ & 0.36     & $\times$ & 0.43     & $\times$ & 0.27 & $\times$ \\
TBCL & 0.38     & 0.63     & 0.38     & $\times$ & $\times$ & $\times$ & 0.26     & $\times$ & 0.42     & $\times$ & 0.24 & $\times$ \\
TBCD & 0.38     & $\times$ & 0.31     & $\times$ & 0.27     & $\times$ & 0.23     & $\times$ & 0.36     & $\times$ & 0.26 & $\times$ \\
\hline
Lips    & 0.37 & $\times$  & 0.39     & 0.41     & 0.35     & 0.57     & 0.26     & $\times$ & 0.25     & $\times$ & 0.39 & $\times$ \\
Tongue  & 0.44 & 0.57      & 0.35     & 0.44     & $\times$ & $\times$ & 0.34     & $\times$ & 0.40     & $\times$ & 0.23 & $\times$ \\
Overall & 0.47 & 0.52      & 0.43     & 0.57     & 0.30     & 0.45     & 0.33     & $\times$ & 0.38     & $\times$ & 0.38 & $\times$ \\
\bottomrule
\end{tabular}
\end{table*}

\section{Classifying Depressed vs Control Speakers}
\label{Classification Experiments}
This section presents the feasibility of using the proposed biomarkers as features in a Machine Learning system to distinguish between the depressed and control speakers. To evaluate this, the three biomarkers (LLE, CD and SE) are computed for each of the six TVs, yielding a feature vector of dimension 18 corresponding to each audio sample. These features are used as input to a Support Vector Machine (SVM). Table \ref{tab:classification} reports the classification results achieved using a 5-fold cross-validation in terms of accuracy, precision, recall and F1. The results show that the biomarkers can achieve better results for the Reading task when compared to a similar classification baseline: Low Level Descriptors (LLDs) from speech input to an SVM as reported by \citet{Tao2023}. The results are comparable for the Interview task. This result is encouraging as it demonstrates that the biomarkers are powerful enough to distinguish between the two classes.
\begin{table*}[tb]
\centering
\caption{Classification performance of the proposed biomarkers compared with the Low Level Descriptors (LLDs) baseline~\cite{Tao2023} on the Androids Corpus for the Reading and Interview tasks. Results are reported as mean(±standard deviation). Best values per metric are highlighted in bold. The sign (*) denotes statistically significant differences compared to the random classifier.}
\label{tab:classification}
\begin{tabular*}{\textwidth}{@{\extracolsep{\fill}}lcccc}
\hline
 & \textbf{Accuracy} & \textbf{Precision} & \textbf{Recall} & \textbf{F1-score} \\
\hline
Random Classifier 
& 50.1 & 51.8 & 51.8 & 51.8 \\
\hline
\multicolumn{5}{c}{\textbf{Reading Task}} \\
\hline
LLDs + SVM \cite{Tao2023}
& 69.6*$\pm$5.3 & \textbf{73.6*$\pm$19.1} & 68.8*$\pm$12.0 & 68.4*$\pm$7.7 \\
(LLE, CD, SE) + SVM
&\textbf{ 73.2*$\pm$7.44} & 71.73*$\pm$16.67 & \textbf{70.12*$\pm$13.21 }& \textbf{70.44*$\pm$13.31} \\
\hline
\multicolumn{5}{c}{\textbf{Interview Task}} \\
\hline
LLDs + SVM~\cite{Tao2023}
& \textbf{73.3*$\pm$10.6 }& \textbf{73.5*$\pm$16.1} & \textbf{74.5*$\pm$13.2 }& \textbf{73.6*$\pm$13.6} \\
(LLE, CD, SE) + SVM 
& 68.16*$\pm$3.2 & 73.15*$\pm$15.32 & 62.99$\pm$13.38 & 65.61*$\pm$7.79 \\
\hline
\end{tabular*}
\end{table*}

\section{Conclusions}\label{concl}
This article proposed new depression biomarkers in speech, namely LLE, CD and SE, that are derived from the dynamics of the TVs in speech articulation. LLE  is seen as a measure of predictability, CD measures complexity and the SE is an information-theoretic measure of randomness. Unlike past studies that used self-reported depression scores, this work used our own Androids Corpus~\cite{Tao2023}, involving 64 speakers diagnosed with depression by professional clinicians and 54 control speakers with no history of mental health issues. All three biomarkers show significant differences between the depressed and the control participants. To the best of our knowledge, this is the first work showing that such measurements are effective depression biomarkers.

The experiments considered read (reading a piece of text) and spontaneous speech (collected through interviews), and thus make it possible to analyze the effectiveness of the biomarkers both in the presence and absence of the cognitive effort required to plan what to say next - an important factor in detecting depressive symptoms~\cite{Alpert2001}. The only marker that appears to be affected by the added cognitive load during interviews is the LLE, which is effective only for read speech. This suggests that the cognitive effort involved in spontaneous speech might make the TV dynamics less predictable in general, thereby reducing any observable difference between depressed and control speakers. The other two measurements (CD and SE) are equally effective for both read and spontaneous speech and are not affected by the changes in cognitive load. Most likely, the differences observed for CD and SE depend rather on psychomotor retardation, i.e., on the influence of depression on motor skills~\cite{Caligiuri2000}. 

The main limitation of this work, and more generally of the research on biomarkers, is that these are supposed to be objective traces of depression, but their definition depends on the judgment of the clinicians. Any measurable aspect of speech is considered an effective biomarker whenever it is consistent with clinical diagnoses that, although rigorous, could still be biased and subjective~\cite{Koops2023}.  
On the other hand, in the case of the Androids Corpus, the patients were diagnosed before the collection of the data and they interact face-to-face with clinicians they have been meeting for several years. Therefore, the diagnosis does not result only from speech data, but from all behavioural channels available during co-located meetings (facial expressions, words, posture, etc.), including all the information recorded about patients in the healthcare centre. In this respect, the results of Section~\ref{expres} are unlikely to depend on a mere tendency of the clinicians to diagnose people who speak in a certain way as depressed.

Also note that the number of female speakers is twice the number of male speakers (see Section~\ref{expres}). This is expected as all epidemiological studies show that women tend to develop depression more frequently than men~\cite{Kuehner2017}. Consequently, when analyzing male and female speakers separately, the effects are greater for females. However, when limiting the female speakers to the number of available male speakers (32 persons), the performance of the biomarkers is comparable across genders. This suggests that the observed differences do not depend on gender, but on the number of available samples for each gender.

The experiments consider every TV individually, but the strongest effects, as measured by the absolute value of the Cliff's $\delta$, are observed for the TV groups corresponding to lips and tongue regions or to the whole set of TVs. The highest value of $|\delta|$ for an individual TV is 0.41 (TBCL in the case of LLE for read speech), while it is 0.51 for a group of TVs (tongue in the case of LLE for read speech). Furthermore, $|\delta|\ge 0.4$ for only 1 individual TV out of 36 comparisons between depressed and control speakers, while $|\delta|>0.4$ 6 times out of 18 in the case of TV groups. This is not surprising because the pellets do not move independently, but in coordination, at least at the level of distinct organs such as tongue and lips. In other words, the discrimination power of a biomarker improves when considering the value of a TV in the ``context'' of the other TVs and not in isolation.  

Overall, the results in this work suggest that, in depressed speakers, articulation is less predictable (greater LLE), more constrained (lower CD), and more repetitive (lower SE). Although they do not necessarily correspond to any specific perceptual characteristics of speech (e.g., lower loudness or slower speaking rate), the biomarkers account for the presence of the pathology consistently and reliably. Future work will focus on using them as a representation for more sophisticated automatic depression detection systems, possibly in conjunction with foundation models.
More broadly, using TVs as observables of a dynamical system opens new ways to explore aspects of speech pathology that, to the best of our knowledge, have not been investigated so far.

%


\section{Author Declarations} The authors of this work have no Conflicts of Interest.

\section{Data Availability Statement}
All the experiments of this work have been performed over the Androids Corpus~\cite{Tao2023}, a publicly available depression detection benchmark (see footnote~\ref{androidsfootnote} for the download link).

\begin{acknowledgments}
Alessandro Vinciarelli is supported by the UKRI through the UKRI Centre for Doctoral Training in Socially Intelligent Artificial Agents (EP/S02266X/1).
\end{acknowledgments}
%





\bibliographystyle{jasaauthyear2}
\bibliography{references}




\end{document}